\documentclass[12pt,onecolumn,journal]{IEEEtran}
\usepackage{cite}
\usepackage{graphicx}
\usepackage{psfrag}
\usepackage{epsfig}
\usepackage{subfigure}
\graphicspath{{../eps/}{../ps/}}

\begin{document}

\title{Capacity and Performance of Adaptive MIMO System Based on Beam-Nulling}

\author{Mabruk~Gheryani, Zhiyuan~Wu, and Yousef~R.~Shayan\\Concordia University, Department of Electrical Engineering \\Montreal,
Quebec, H4G 2W1, Canada \\email: (m\_gherya, zy\_wu,
yshayan)@ece.concordia.ca}

\maketitle
\begin{abstract}
In this paper, we propose a scheme called ``beam-nulling" for MIMO
adaptation. In the beam-nulling scheme, the eigenvector of the
weakest subchannel is fed back and then signals are sent over a
generated subspace orthogonal to the weakest subchannel.
Theoretical analysis and numerical results show that the capacity
of beam-nulling is closed to the optimal water-filling at medium
SNR. Additionally, signal-to-interference-plus-noise ratio (SINR)
of MMSE receiver is derived for beam-nulling. Then the paper
presents the associated average bit-error rate (BER) of
beam-nulling numerically which is verified by simulation.
Simulation results are also provided to compare beam-nulling with
beamforming. To improve performance further, beam-nulling is
concatenated with linear dispersion code. Simulation results are
also provided to compare the concatenated beam-nulling scheme with
the beamforming scheme at the same data rate. Additionally, the
existing beamforming and new proposed beam-nulling can be extended
if more than one eigenvector is available at the transmitter.  The
new extended schemes are called multi-dimensional (MD) beamforming
and MD beam-nulling. Theoretical analysis and numerical results in
terms of capacity are also provided to evaluate the new extended
schemes. Simulation results show that the MD scheme with LDC can
outperform the MD scheme with STBC significantly when the data
rate is high.
\end{abstract}


%
\IEEEpeerreviewmaketitle

\section{Introduction}
Since the discovery of multiple-input-multiple-output (MIMO)
capacity\cite{Telatar:MIMOCapacity}\cite{Foschini:MIMOCapacity}, a
lot of research efforts have been put into this field. It has been
recognized that adaptive techniques proposed for
single-input-single-output (SISO)
channels\cite{Cavers:VR_transmission}\cite{Goldsmith:VR_MQAM} can
also be applied to improve MIMO channel capacity.

The ideal scenario is that the transmitter has full knowledge of
channel state information (CSI). Given this perfect CSI feedback,
the original MIMO channel can be converted to multiple uncoupled
SISO channels via singular value decomposition (SVD) at the
transmitter and the receiver\cite{Telatar:MIMOCapacity}. In other
words, the original MIMO channel can be decomposed into several
orthogonal ``spatial subchannels" with various propagation gains.

To achieve better performance, various schemes can be implemented
depending on the availability of CSI at the transmitter
\cite{Luo:Capacity_Time-Varying}-\cite{Giannakis:Multiantenna-Beamforming-Constrained-Feedback}.
 If the transmitter has full knowledge of the channel matrix, i.e., full CSI, the
so-called ``water-filling" (WF) principle is performed on each
spatial subchannel to maximize the channel capacity
\cite{Telatar:MIMOCapacity}. This scheme is optimal in this case.
Various WF-based schemes have been proposed, such as
\cite{Shen:Comparison_Water-filling}\cite{Zhang:QoS_WF}. For the
WF-based scheme, the feedback bandwidth for the full CSI grows
with respect to the number of transmit and receive antennas and
the performance is often very sensitive to channel estimation
errors.

To mitigate these disadvantages, various beamforming (BF)
techniques for MIMO channels have also been investigated
intensively. In an adaptive beamforming scheme, the complex
weights of the transmit antennas are fed back from the receiver.
If only one eigenvector can be fed back, eigen-beamforming
\cite{Giannakis:Optimal-STBC-Channel-Mean} is optimal. The
eigen-beamforming scheme only applies to the strongest spatial
subchannel but can achieve full diversity and high signal-to-noise
ratio (SNR) \cite{Giannakis:Optimal-STBC-Channel-Mean}. Also, in
practice, the eigen-beamforming scheme has to cooperate with the
other adaptive parameters to improve performance and/or data rates
such as constellation and coding rate. There are also other
beamforming schemes based on various criteria. Examples of such
schemes are \cite{Giannakis:Optimal-STBC-Channel-Mean}
-\cite{Cavers:BF}. Note that the conventional beamforming is
optimal in terms of maximizing the SNR at the receiver. However,
it is sub-optimal from the MIMO capacity perspective, since only a
single data stream, as opposed to parallel streams, is transmitted
through the MIMO channel \cite{Zhou:Channel_Prediction}.

In this paper, we propose a new technique called ``beam-nulling"
(BN). This scheme uses the same feedback bandwidth as beamforming,
that is, only one eigenvector is fed back to the transmitter. The
beam-nulling transmitter is informed by the weakest spatial
subchannel and, where both transmitter and receiver know how to
generate the same spatial subspace, sends signals over a generated
spatial subspace orthogonal to the weakest subchannel. Although
the transmitted symbols are ``precoded" according to the feedback,
beam-nulling is different from the other existing precoding
schemes with limited feedback channel, which are independent of
the instantaneous channel but the optimal precoding depends on the
instantaneous channel
\cite{Love:Beamforming}\cite{Zheng:Capacity_Precode}.

Using this new techniques instead of the optimal water-filling
scheme, the loss of channel capacity can be reduced. This paper
also addresses the performance of beam-nulling. To achieve better
performance, beam-nulling can be concatenated with the other
space-time (ST) coding schemes, such as space-time trellis codes
(STTCs)\cite{Tarokh:STC_Perf_Const}, space-time block codes
(STBCs)\cite{Alamouti:STBC}\cite{Tarokh:STBC} and linear
dispersion codes (LDCs)\cite{Hassibi:LDC}-\cite{Wu:Design4CSTM},
etc. For simplicity and flexibility, LDCs are preferable. We
provide numerical and simulation results are provided to
demonstrate the merits of the new proposed scheme. Additionally,
if more than one eigenvector, e.g. $k$ eigenvectors, can be
available at the transmitter, the existing beamforming scheme and
the proposed beam-nulling scheme can be further extended,
respectively. The extended schemes will exploit or discard $k$
spatial subchannels and they will be referred to as
``multi-dimensional (MD)" beamforming and ``multi-dimensional"
beam-nulling, respectively.

This paper will be organized as follows. Our channel model is
presented in Section~\ref{sec_sys_mod}. In Section~\ref{sec_PA},
four power allocation strategies, i.e., equal power,
water-filling, eigen-beamforming, and a new power allocation
strategy called ``beam-nulling" are studied and compared in terms
of channel capacity. In Section \ref{sec_perf_BN}, bit error rate
(BER) of the proposed beam-nulling scheme using MMSE detector is
studied and verified. The proposed scheme is compared with the
eigen-beamforming scheme at various data rates in terms of BER.
Beam-nulling concatenated with LDC is proposed and evaluated. In
Section \ref{sec_ext_frame}, extended adaptive frameworks, i.e.,
MD beamforming and MD beam-nulling, are proposed. Capacity and
performance of these two schemes are discussed and compared. To
improve performance further and maintain reasonable complexity, MD
schemes concatenated with linear space-time codes, such as STBC
and LDC, are proposed and evaluated. Finally, in
Section~\ref{sec_con}, conclusions are drawn.

\section{Channel Model}\label{sec_sys_mod}
In this study, the channel is assumed to be a Rayleigh flat fading
channel with $N_t$ transmit and $N_r$ ($N_r \geq N_t$) receive
antennas. We denote the complex gain from the transmit antenna $n$
to the receiver antenna $m$ by $h_{mn}$ and collect them to form
an $N_r \times N_t$ channel matrix $\mathbf{H}=[h_{mn}]$. The
channel is known perfectly at the receiver. The entries in
$\mathbf{H}$ are assumed to be independent and identically
distributed (\emph{i.i.d.}) symmetrical complex Gaussian random
variables with zero mean and unit variance.

The symbol vector at the $N_t$ transmit antennas is denoted by
$\mathbf{x}=[x_1, x_2, \ldots, x_{N_t}]^T$. According to
information theory \cite{Shannon:info_theory}, the optimal
distribution of the transmitted symbols is Gaussian. Thus, the
elements $\{x_i\}$ of $\mathbf{x}$ are assumed to be \emph{i.i.d.}
Gaussian variables with zero mean and unit variance, i.e.,
$E(x_i)=0$ and $E|x_i|^2=1$.

The singular-value decomposition of $\mathbf{H}$ can be written as
\begin{equation}\label{svd_H}
\mathbf{H}=\mathbf{U}\mathbf{\Lambda}\mathbf{V}^{H}
\end{equation}
where $\mathbf{U}$ is an $N_r \times N_r$ unitary matrix,
$\mathbf{\Lambda}$ is an $N_r \times N_t$ matrix with singular
values $\{\lambda_{i}\}$ on the diagonal and zeros off the
diagonal, and $\mathbf{V}$ is an $N_t \times N_t$ unitary matrix.
For convenience, we assume $\lambda_{1}\geq \lambda_{2} \ldots
\geq \lambda_{N_t}$, $\mathbf{U}=[\mathbf{u}_1 \mathbf{u}_2 \ldots
\mathbf{u}_{N_r} ]$ and $\mathbf{V}=[\mathbf{v}_1 \mathbf{v}_2
\ldots \mathbf{v}_{N_t} ]$. $\{\mathbf{u}_i\}$ and $\mathbf{v}_i$
are column vectors.
From equation (\ref{svd_H}), the original
channel can be considered as consisting of uncoupled parallel
subchannels. Each subchannel corresponds to a singular value of
$\mathbf{H}$. In the following context, the subchannel is also
referred to as ``spatial subchannel". For instance, one spatial
subchannel corresponds to $\lambda_{i}$, $\mathbf{u}_i$ and
$\{\mathbf{v}_i\}$.

\section{Power Allocation Among Spatial Subchannels}\label{sec_PA}

We assume that the total transmitted power is constrained to $P$.
Given the power constraint, different power allocation among
spatial subchannels can affect the channel capacity tremendously.
Depending on power allocation strategy among spatial subchannels,
four schemes are presented which are equal power, water-filling,
eigen-beamforming, and the new power allocation which is
beam-nulling.

If the transmitter has no knowledge about the channel, the most
judicious strategy is to allocate the power to each transmit
antenna equally, i.e., equal power. In this case, the received
signals can be written as
\begin{equation}\label{sys_mod4eq}
\mathbf{y}=\sqrt{\frac{P}{N_t}}\mathbf{H}\mathbf{x}+\mathbf{z}
\end{equation}
\(\mathbf{z}\) is the additive white Gaussian noise (AWGN) vector
with \emph{i.i.d.} symmetrical complex Gaussian elements of zero
mean and variance \(\sigma_z^2\). The associated ergodic channel
capacity can be written as \cite{Telatar:MIMOCapacity}
\begin{equation}\label{cap_equal}
\bar{C}_{eq}=E\left[\sum\limits_{i=1}^{N_t}\log\left(1+\frac{\rho}{N_t
}\lambda_{i}^2\right)\right]
\end{equation}
where $E[\cdot]$ denotes expectation with respect to $\mathbf{H}$
and $\rho=\frac{P}{\sigma^{2}_{z}}$ denotes SNR.
 If the transmitter has full knowledge about the channel, the most judicious strategy
is to allocate the power to each spatial subchannel by
water-filling principle \cite{Telatar:MIMOCapacity}. In
water-filling scheme, the received signals can be written as
\begin{equation}\label{WF_subchan}
\tilde{y}_i=\sqrt{P_i} \lambda_{i} x_i+\tilde{z}_i
\end{equation}
where $\sum\limits_{i=1}^{N_t}P_i=P$ as a constraint and
$\tilde{z}_i$ is the AWGN random variable with zero mean and
$\sigma_z^2$ variance. Following the method of Lagrange
multipliers, $P_i$ can be found \cite{Telatar:MIMOCapacity} and
the total ergodic channel capacity is
\begin{equation}\label{cap_wf}
\bar{C}_{wf}=E\left[\sum\limits_{i=1}^{N_t}\log\left(1+\frac{P_i}{\sigma^{2}_{z}}\lambda_{i}^2\right)\right]
\end{equation}

To save feedback bandwidth, beamforming can be considered. For the
MIMO model, the optimal beamforming is called ``eigen-beamforming"
\cite{Giannakis:Optimal-STBC-Channel-Mean}, or simply beamforming.
We assume one symbol, saying $x_1$, is transmitted. At the
receiver, the received vector can be written as
\begin{equation}\label{eigen_rec_y}
\mathbf{y}_1=\sqrt{P} \mathbf{H}\mathbf{v}_{1}x_1+\mathbf{z}_1
\end{equation}
where $\mathbf{z}_1$ is the additive white Gaussian noise vector
with \emph{i.i.d.} symmetrical complex Gaussian elements of zero
mean and variance $\sigma^{2}_{z}$. The associated ergodic channel
capacity can be written as
\begin{equation}\label{cap_bf}
\bar{C}_{bf}=E\left[\log\left(1+\rho\lambda_{1}^2 \right)\right]
\end{equation}

The eigen-beamforming scheme can save feedback bandwidth and is
optimized in terms of SNR \cite{Cavers:BF}. However, since only
one spatial subchannel is considered, this scheme suffers from
loss of channel capacity \cite{Zhou:Channel_Prediction},
especially when the number of antennas grows.

\subsection{Beam-Nulling}\label{subsec_BN_model}

The eigen-beamforming scheme can save feedback bandwidth and is
optimized in terms of SNR \cite{Cavers:BF}. However, since only a
single spatial subchannel is considered, this scheme suffers from
loss of channel capacity \cite{Zhou:Channel_Prediction},
especially when the number of antennas grows. Inspired by the
eigen-beamforming scheme, we will propose a new beamforming-like
scheme called ``beam-nulling" (BN). This scheme uses the same
feedback bandwidth as beamforming, that is, only one eigenvector
is fed back to the transmitter. Unlike the eigen-beamforming
scheme in which only the best spatial subchannel is considered,
the beam-nulling scheme discards only the worst spatial
subchannel. Hence, in comparison with the optimal water-filling
scheme, the loss of channel capacity can be reduced.

\begin{figure}
\centering
\includegraphics[width=3in]{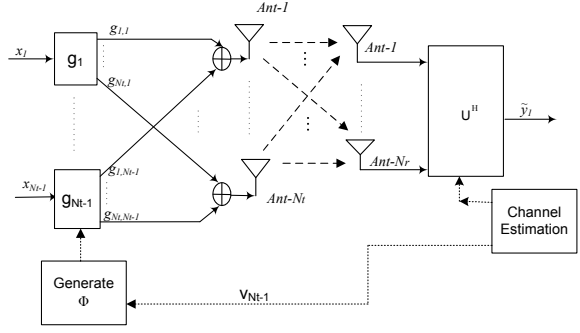}
\caption{\label{fig:1D_BN_scheme} beam-nulling scheme.}
\end{figure}

In this scheme as shown in Fig.~\ref{fig:1D_BN_scheme}, the
eigenvector associated with the minimum singular value from the
transmitter side, i.e., $\mathbf{v}_{N_t}$, is fed back to the
transmitter. A subspace orthogonal to the weakest spatial channel
is constructed so that the following condition is satisfied.
\begin{equation}\label{orth_subspace}
\mathbf{\Phi}^{H}\mathbf{v}_{N_t}=\mathbf{0}
\end{equation}
The $N_t \times (N_t-1)$ matrix $\mathbf{\Phi}=[\mathbf{g}_1
\mathbf{g}_2 \ldots \mathbf{g}_{N_t-1}]$ spans the subspace. Note
that the method to construct the subspace $\mathbf{\Phi}$ should
also be known to the receiver.

Here is an example of construction of the orthogonal subspace. We
construct an $N_t \times N_t$ matrix
\begin{equation}\label{A_construction}
\mathbf{A}=[\mathbf{v}_{N_t} \mathbf{I}']
\end{equation}
where $\mathbf{I}'=[\mathbf{I}_{(N_t-1)\times(N_t-1)}
\mathbf{0}_{(N_t-1)\times 1}]^T$. Applying QR decomposition to
$\mathbf{A}$, we have
\begin{equation}\label{QR_Decom}
\mathbf{A}=[\mathbf{v}_{N_t} \mathbf{\Phi}]\cdot\mathbf{\Gamma}
\end{equation}
where $\mathbf{\Gamma}$ is an upper triangular matrix with the
(1,1)-th entry equal to 1. $\mathbf{\Phi}$ is the subspace
orthogonal to $\mathbf{v}_{N_t}$.

At the transmitter, $N_t-1$ symbols denoted as $\mathbf{x}'$ are
transmitted over the orthogonal subspace $\mathbf{\Phi}$. The
received signals at the receiver can be written as

\begin{eqnarray}\label{BN_rec_y}
\nonumber
  \mathbf{y}' &=& \sqrt{\frac{P}{N_t-1}}\mathbf{H}\mathbf{\Phi}\mathbf{x}'+\mathbf{z}' \\
   &=& \widehat{\mathbf{H}}\mathbf{x}'+\mathbf{z}'
\end{eqnarray}
where $\mathbf{z}'$ is additive white Gaussian noise vector with
\emph{i.i.d.} symmetrical complex Gaussian elements of zero mean
and variance $\sigma^{2}_{z}$ and
$\widehat{\mathbf{H}}=\sqrt{\frac{P}{N_t-1}}\mathbf{H}\mathbf{\Phi}$.

Substituting (\ref{svd_H}) into (\ref{BN_rec_y}) and multiplying
$\mathbf{y}'$ by $\mathbf{U}^{H}$, results in
\begin{equation}\label{BN_tld_y}
\widetilde{\mathbf{y}}=\sqrt{\frac{P}{N_t-1}}\mathbf{\Lambda}\left(%
\begin{array}{c}
  \mathbf{B} \\
  \mathbf{0}^{T} \\
\end{array}%
\right)\mathbf{x}'+\widetilde{\mathbf{z}}
\end{equation}
where $\widetilde{\mathbf{z}}$ is additive white Gaussian noise
vector with \emph{i.i.d.} symmetrical complex Gaussian elements of
zero mean and variance $\sigma^{2}_{z}$. With the condition in
(\ref{orth_subspace}),
\begin{equation}
\mathbf{V}^{H}\mathbf{\Phi}=\left(%
\begin{array}{c}
  \mathbf{B} \\
  \mathbf{0}^{T} \\
\end{array}%
\right)\end{equation} where
\begin{equation}\label{matrix_B}
\mathbf{B}=\left(%
\begin{array}{cccc}
  \mathbf{v}_{1}^{H}\mathbf{g}_{1} & \mathbf{v}_{1}^{H}\mathbf{g}_{2} & \ldots & \mathbf{v}_{1}^{H}\mathbf{g}_{N_t-1} \\
  \mathbf{v}_{2}^{H}\mathbf{g}_{1} & \ddots & \ldots & \vdots \\
  \vdots & \vdots & \ddots & \vdots \\
  \mathbf{v}_{N_t-1}^{H}\mathbf{g}_{1}& \ldots & \ldots & \mathbf{v}_{N_t-1}^{H}\mathbf{g}_{N_t-1} \\
\end{array}%
\right)
\end{equation}

$\mathbf{B}$ is an $(N_t-1)\times(N_t-1)$ unitary matrix. From
(\ref{BN_tld_y}), the available spatial channels are $N_t-1$.
Since the weakest spatial subchannel is ``nulled" in this scheme,
power can be allocated equally among the other $N_t-1$
subchannels. Equation (\ref{BN_tld_y}) can be rewritten as
\begin{equation}\label{BN_tld_y_1}
\widetilde{\mathbf{y}}'=\sqrt{\frac{P}{N_t-1}}\mathbf{\Lambda}'\mathbf{B}\mathbf{x}'+\widetilde{\mathbf{z}}'
\end{equation}
where $\widetilde{\mathbf{y}}'$ and $\widetilde{\mathbf{z}}'$ are
column vectors with the first $(N_r-1)$ elements of
$\widetilde{\mathbf{y}}$ and $\widetilde{\mathbf{z}}$,
respectively, and $\mathbf{\Lambda}'=\mathrm{diag}[\lambda_{1},
\lambda_{2}, \ldots, \lambda_{(N_t-1)}]$. From (\ref{BN_tld_y_1}),
the associated ergodic channel capacity can be found as
\begin{equation}\label{cap_bn}
\bar{C}_{bn}=E\left[\sum\limits_{i=1}^{N_t-1}\log\left(1+\frac{\rho}{N_t-1}\lambda_{i}^2\right)\right]
\end{equation}

As can be seen, the beam-nulling scheme only needs one eigenvector
to be fed back. However, since only the worst spatial subchannel
is discarded, this scheme can increase channel capacity
significantly as compared to the conventional beamforming scheme.

\subsection{Comparisons Among the Four Schemes}

In this section, we compare the new proposed beam-nulling scheme
with the other schemes, i.e., equal power, beamforming and
water-filling schemes. Water-filling is the optimal solution among
the four schemes for any SNR.

Differentiating the above ergodic capacities with respect to
$\rho$ respectively, we have
\begin{eqnarray}
  \frac{\partial \bar{C}_{eq}}{\partial \rho} &=& E\left[\sum\limits_{i=1}^{N_t}\frac{1}{\rho+\frac{N_t}{\lambda_{i}^2}}\right] \\
  \frac{\partial \bar{C}_{bf}}{\partial \rho} &=& E\left[\frac{1}{\rho+\frac{1}{\lambda_{1}^2}}\right] \\
  \frac{\partial \bar{C}_{bn}}{\partial \rho} &=& E\left[\sum\limits_{i=1}^{N_t-1}\frac{1}{\rho+\frac{N_t-1}{\lambda_{i}^2}}\right] \\\nonumber
\end{eqnarray}
The differential will also be referred to as ``slope". Since the
second order differentials are negative, the above ergodic
capacities are concave and monotonically increasing with respect
to $\rho$.

With the fact that $\lambda_{1} \geq \lambda_{2} \ldots \geq
\lambda_{N_t}$, it can be readily checked that the slopes of
ergodic capacities associate with equal power and beam-nulling are
bounded as follows.
\begin{eqnarray}
  E\left(\frac{N_t}{\rho+\frac{N_t}{\lambda_{1}}}\right) \geq \frac{\partial \bar{C}_{eq}}{\partial \rho} \geq E\left(\frac{N_t}{\rho+\frac{N_t}{\lambda_{N_t}}}\right)\\
  E\left(\frac{N_t-1}{\rho+\frac{N_t-1}{\lambda_{1}}}\right) \geq \frac{\partial \bar{C}_{bn}}{\partial \rho} \geq E\left(\frac{N_t-1}{\rho+\frac{N_t-1}{\lambda_{(N_t-1)}}}\right)\\\nonumber
\end{eqnarray}

For the case of $N_t=2$, beamforming and beam-nulling have the
same capacity for any  $\rho$ as can be seen from equations of
capacity and slope. If $\rho \rightarrow 0$, equivalently at low
SNR, it can be easily found that
\begin{equation}\label{diff_low_SNR}
\frac{\partial \bar{C}_{bf}}{\partial \rho} \geq \frac{\partial
\bar{C}_{bn}}{\partial \rho} \geq \frac{\partial
\bar{C}_{eq}}{\partial \rho}, \rho \rightarrow 0
\end{equation}
If $\rho \rightarrow \infty$, equivalently at high SNR, it can be
easily found that
\begin{equation}\label{diff_high_SNR}
\frac{\partial \bar{C}_{eq}}{\partial \rho} \geq \frac{\partial
\bar{C}_{bn}}{\partial \rho} \geq \frac{\partial
\bar{C}_{bf}}{\partial \rho}, \rho \rightarrow \infty
\end{equation}
Note that $\bar{C}_{bf}=\bar{C}_{bn}=\bar{C}_{eq}=0$ when $\rho=0$
or minus infinity in dB. Hence, at medium SNR, $\frac{\partial
\bar{C}_{bn}}{\partial \rho}$ has the largest value compared to
$\frac{\partial\bar{C}_{bf}}{\partial \rho}$~and~$
\frac{\partial\bar{C}_{eq}}{\partial \rho}$. Therefore, for low,
medium and high SNRs, beamforming, beam-nulling and equal power
have the largest capacities, respectively.

\begin{figure}
\centering
\includegraphics[width=3.25in]{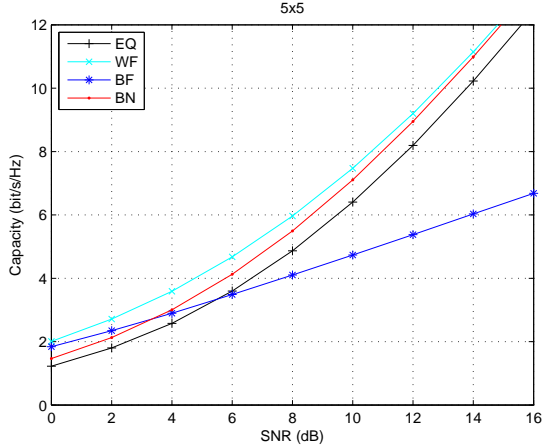}
\caption{\label{5x5_cap}$5 \times 5$ Rayleigh fading channel.}
\end{figure}

In Fig.~\ref{5x5_cap}, capacities of water-filling, beamforming,
beam-nulling and equal power are compared over $5 \times 5$
Rayleigh fading channels, respectively. Note that since SNR is
measured in dB, the curves become convex. In these figures, ``EQ"
stands for equal power, ``WF" stands for water-filling, ``BF"
stands for beamforming and ``BN" stands for beam-nulling. As can
be seen, the water-filling has the best capacity at any SNR
region. The other schemes perform differently at different SNR
regions. At low SNR, the beamforming is the closest to the optimal
water-filling, e.g., the SNR region below $3.5$ dB for $5 \times
5$ fading channel. Note that at low SNR, the water-filling scheme
may only allocate power to one or two spatial subchannels. At
medium SNR, the proposed beam-nulling is the closest to the
optimal water-filling, e.g., the SNR region from $3.5$ dB to $16$
dB for $5 \times 5$ fading channel. The beam-nulling scheme only
discards the weakest spatial subchannel and allocates power to the
other spatial subchannels. As can be seen from the numerical
results, the beam-nulling scheme performs better than the other
schemes in this case. Note that at high SNR, the equal power
scheme will converge with the water-filling scheme.

\section{Performance of Beam-nulling}\label{sec_perf_BN}

\subsection{MMSE Detector}

The close-form error probability for the optimal ML receiver is
difficult to establish. Other suboptimal receivers can also be
implemented. The MMSE detector is especially popular due to its
low complexity and good
performance\cite{Lupas:LMUD}\cite{Poor:prob_MMSE}. In the
following context, BER of the MMSE detector is analyzed for the
beam-nulling scheme.

Let us define
$\widehat{\mathbf{H}}=\sqrt{\frac{P}{N_t-1}}\mathbf{H}\mathbf{\Phi}$
and $\hat{\mathbf{h}}_i$ is the \(i\)-th column of
\(\widehat{\mathbf{H}}\). Equation (\ref{BN_rec_y}) can also be
written as
\begin{equation}\label{sinr_main1_bn}
    \mathbf{y}'=\hat{\mathbf{h}}_i
    x_i+\sum_{j \neq i}\hat{\mathbf{h}}_j x_j+
    \mathbf{z}'
\end{equation}
where $x_i$ is the i-th element of $\mathbf{x}'$.

Without loss of generality, we consider the detection of one
symbol, say \(x_i\). We collect the rest of the symbols into a
column vector \(\mathbf{x}_I\) and denote
$\widehat{\mathbf{H}}_I=[\hat\mathbf{h}_{1},..,
\hat\mathbf{h}_{i-1},\hat\mathbf{h}_{i+1}, ...,
\hat\mathbf{h}_{N_t-1}]$ as the matrix obtained by removing the
\(i\)-th column from \(\widehat{\mathbf{H}}\).

A linear MMSE detector
\cite{Poor:prob_MMSE}\cite{Bohnke:SINR_Analysis} is applied and
the corresponding output is given by
\begin{equation}\label{x_hat_i_bn}
\hat{x}_{i}=\mathbf{w}_{i}^H \mathbf{y}=x_i+ \hat{z}_{i}.
\end{equation}
where \(\hat{z}_{i}\) is the noise term of zero mean.
$\hat{z}_{i}$ can be approximated to be Gaussian
\cite{Poor:prob_MMSE}. The corresponding \(\mathbf{w}_{i}\) can be
found as
\begin{equation}\label{coef_i_bn}
\mathbf{w}_{i}=\frac{\left(\hat{\mathbf{h}}_i
\hat{\mathbf{h}}_i^H+\mathbf{R}_{I}\right)^{-1}
\hat{\mathbf{h}}_i}{\hat{\mathbf{h}}_i^{H}
\left(\hat{\mathbf{h}}_i
\hat{\mathbf{h}}_i^H+\mathbf{R}_{I}\right)^{-1}
\tilde{\mathbf{h}}_i}
\end{equation}

where \(\mathbf{R}_{I}=\widehat{\mathbf{H}}_{I}
\widehat{\mathbf{H}}_{I}^H + \sigma_z^2 \mathbf{I}\). Note that
the scaling factor \(\frac{1}{\hat{\mathbf{h}}_i^{H}
\left(\hat{\mathbf{h}}_i
\hat{\mathbf{h}}_i^H+\mathbf{R}_{I}\right)^{-1}
\hat{\mathbf{h}}_i}\) in the coefficient vector of the MMSE
detector \(\mathbf{w}_{i}\) is added to ensure an unbiased
detection as indicated by (\ref{x_hat_i_bn}). The variance of the
noise term \(\hat{z}_{i}\) can be found from (\ref{x_hat_i_bn})
and (\ref{coef_i_bn}) as
\begin{equation}\label{var_inf_no_bn}
\hat{\sigma}_{i}^{2}=\mathbf{w}_{i}^H \mathbf{R}_{I}
\mathbf{w}_{i}
\end{equation}
Substituting the coefficient vector for the MMSE detector in
(\ref{coef_i_bn}) into (\ref{var_inf_no_bn}), the variance can be
written as
\begin{equation}\label{var_inf_no_mmse_bn}
\hat{\sigma}_{i}^{2}=\frac{1}{\hat{\mathbf{h}}_i^H
\mathbf{R}_{I}^{-1} \hat{\mathbf{h}}_i}
\end{equation}

Then, the SINR of MMSE associated with $x_i$ is
$1/\hat{\sigma}_{i}^{2}$.
\begin{equation}\label{sinr1}
    \gamma_{i}=\frac{1}{\hat{\sigma_{i}^{2}}}
            =\hat{\mathbf{h}}_i^H
            \mathbf{R}_{I}^{-1}
\hat{\mathbf{h}}_i
\end{equation}

The closed-form BER for a channel model such as (\ref{x_hat_i_bn})
can be found in \cite{Proakis:DigComm}. The average BER over MIMO
fading channel for a given constellation can be found for
beam-nulling as follows.
\begin{equation}\label{ber_av_bn_num}
BER_{av}=E_{\gamma_{i}}\left[ \frac{1}{N_t-1} \sum_{i}
BER(\gamma_{i})\right]
\end{equation}

The closed-form formula for the average BER in
(\ref{ber_av_bn_num}) depends on the distribution of $\gamma_{i}$,
which is difficult to determine. Here, the above average BER is
calculated numerically. For example, the average BER for
$2^\eta$-PSK is
\begin{equation}
BER_{av}=E_{\gamma_{i}}\left[ \frac{1}{N_t-1} \sum_{i}
\frac{2}{\eta}Q\left(\sqrt{2
\eta~\gamma_{i}}~\sin(\frac{\pi}{2^{\eta}})\right)\right]
\end{equation}
and the average BER for rectangular $2^{\eta}$-QAM is
\begin{equation}
BER_{av}=E_{\gamma_{i}}\left[ \frac{1}{N_t-1} \sum_{i}
\frac{4}{\eta}Q\left(\sqrt{\frac{3\eta~\gamma_{i}}{2^{\eta}-1}}\right)\right]
\end{equation}
where $Q(\cdot)$ denotes the Gaussian $Q$-function.

\begin{figure}
\centering \subfigure[$3 \times 3$,
8PSK]{\includegraphics[width=3in]{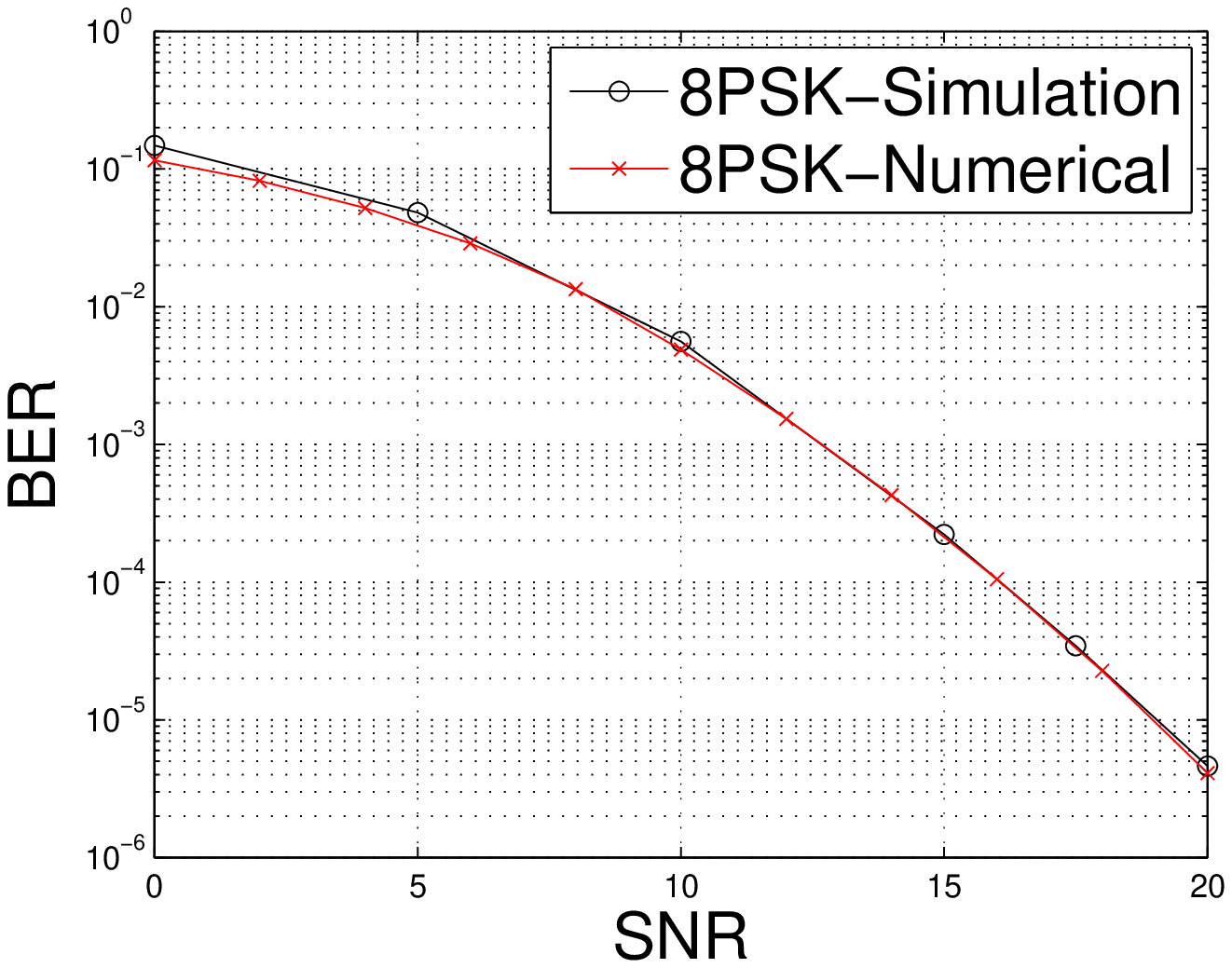}} \hfil
\subfigure[$4 \times 4$,
4PSK]{\includegraphics[width=3in]{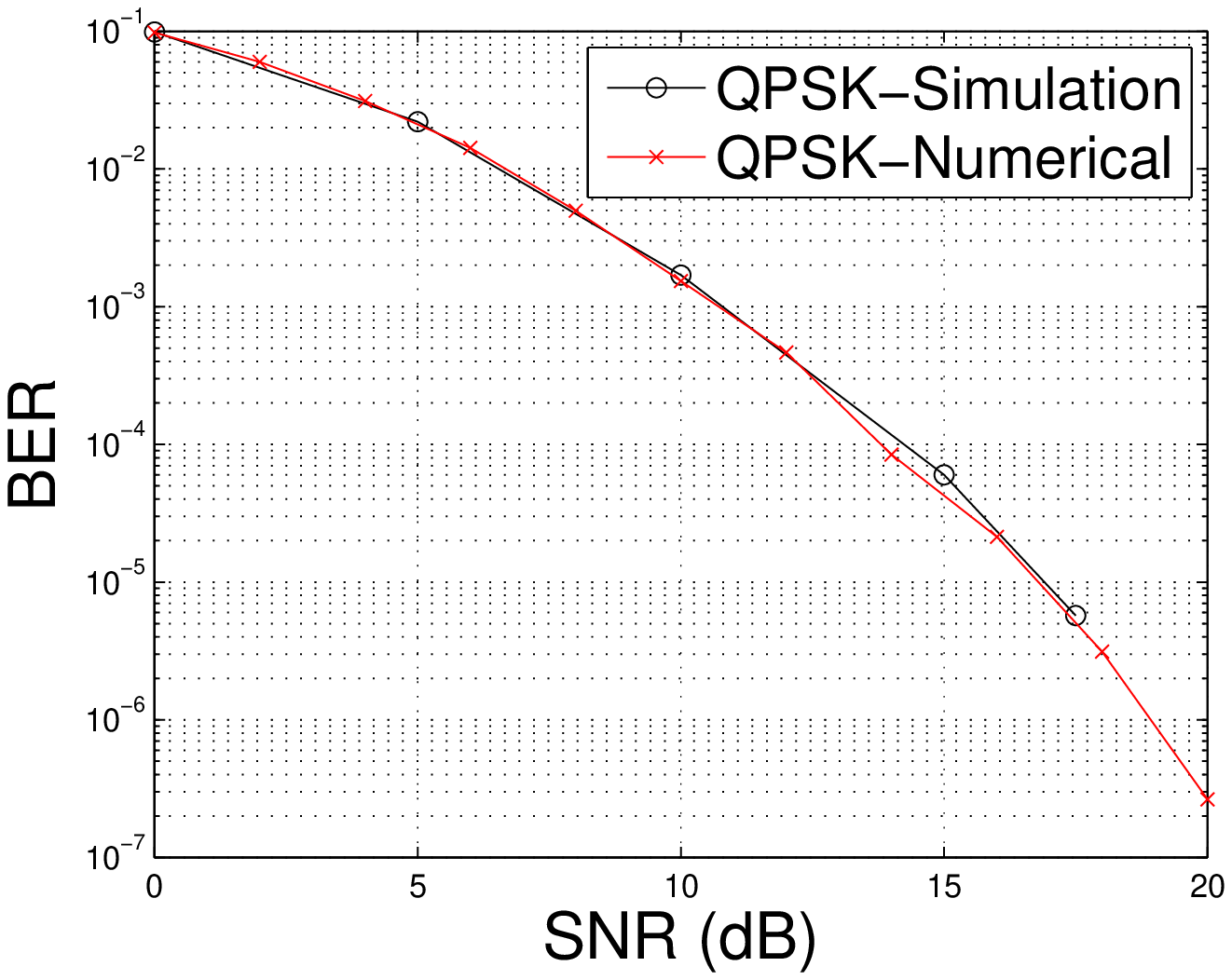}}
\caption{\label{ber_BN_num}Numerical and simulation results for
beam-nulling scheme.}
\end{figure}

In Fig.~\ref{ber_BN_num}, numerical and simulation results are
compared for 8PSK over $3 \times 3$ Rayleigh fading channel and
QPSK over $4 \times 4$ Rayleigh fading channel, respectively. As
can be seen, the numerical and simulation results match well.

\subsection{Performance Comparison Between Beamforming and Beam-nulling}

\begin{figure}
\centering
\subfigure[R=3]{\includegraphics[width=3in]{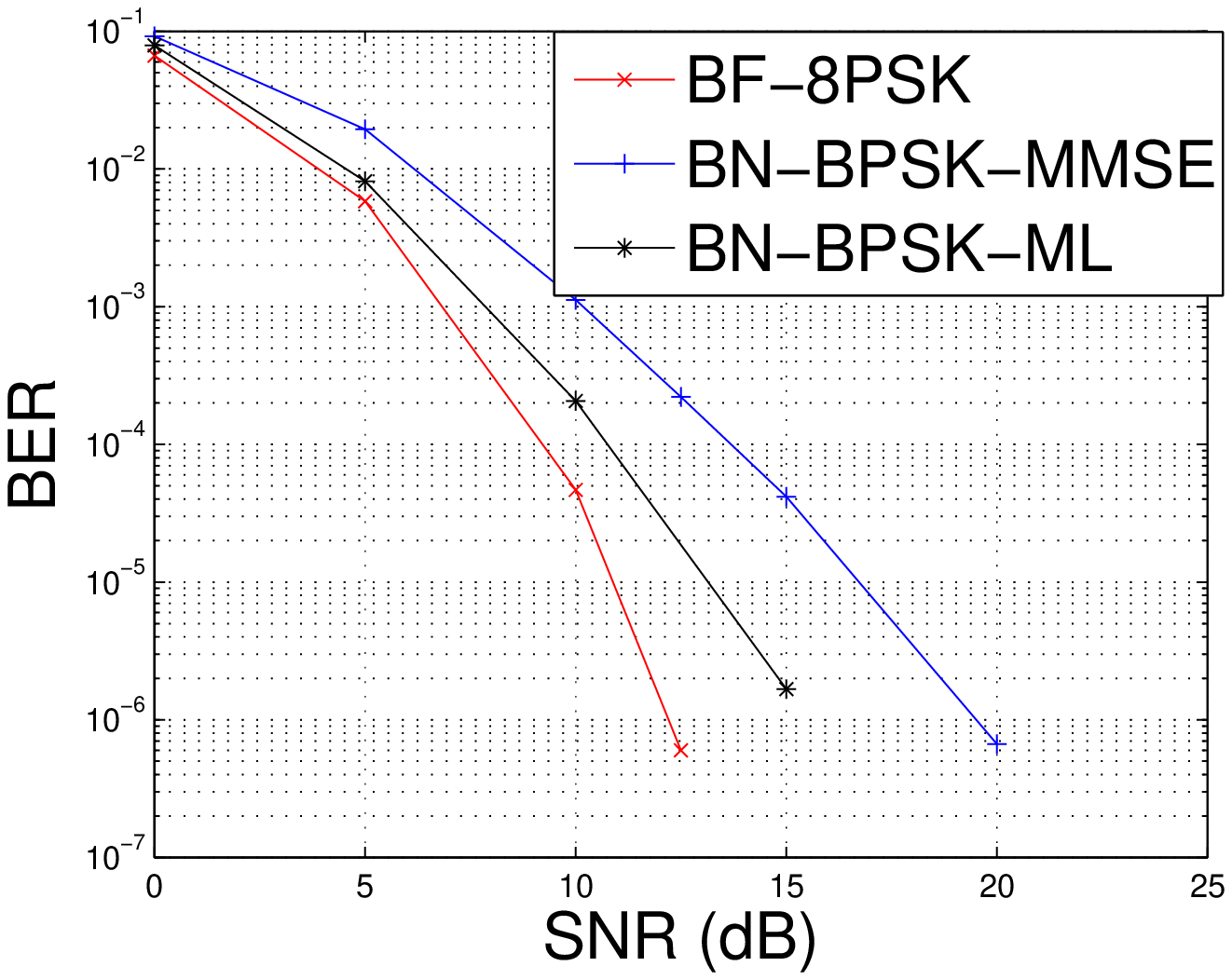}} \hfil
\subfigure[R=6]{\includegraphics[width=3in] {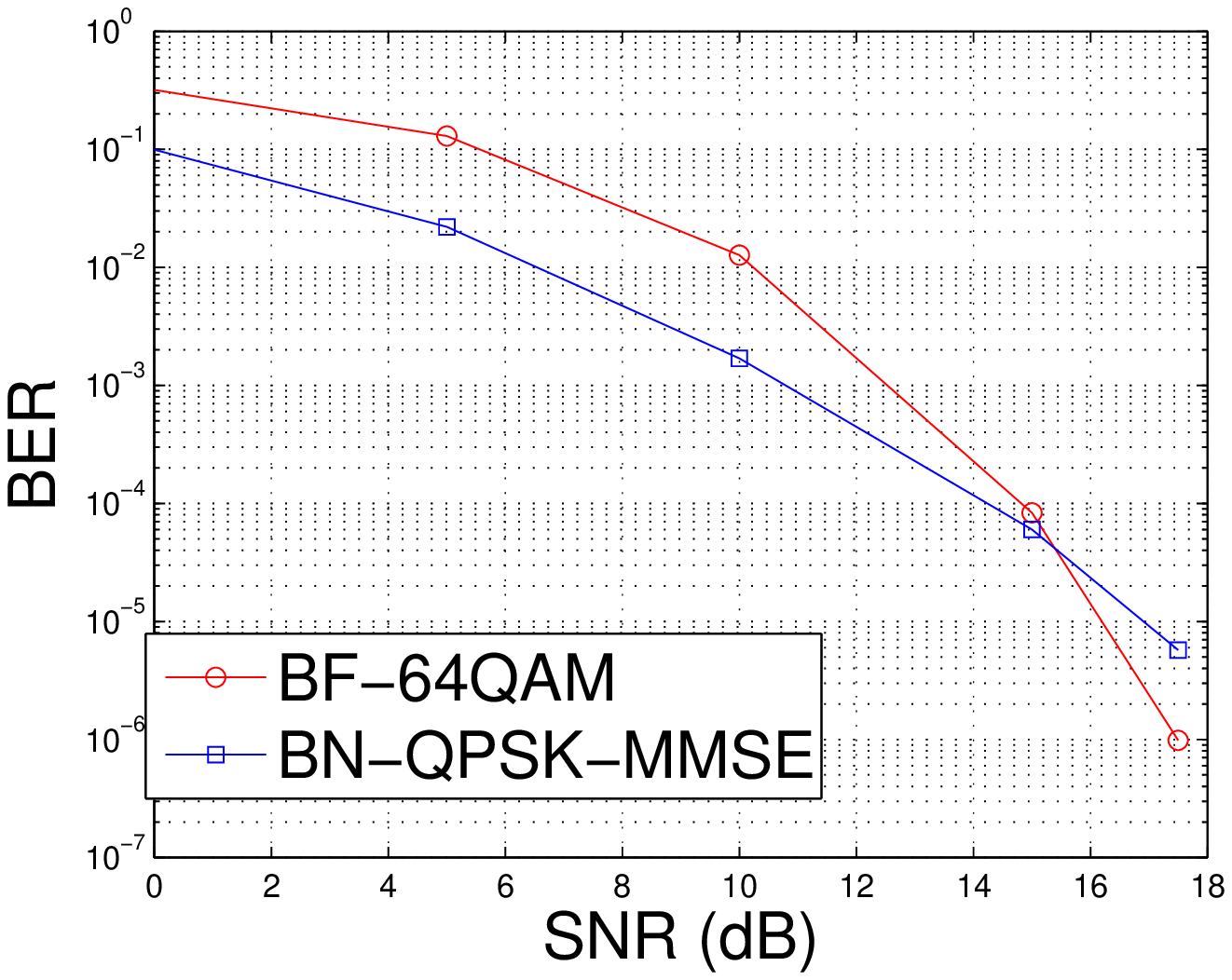}}
\caption{\label{ber_bf_bn_4x4}Comparison over $4 \times 4$
Rayleigh fading channel.}
\end{figure}

In Fig.~\ref{ber_bf_bn_4x4}, simulation results are compared for
various data rates $R$ over $4 \times 4$ Rayleigh fading channels.
In the following simulations, a data rate $R$ is measured in bits
per channel use. The beamforming scheme is equivalent to a SISO
channel using a maximum ratio combining (MRC) receiver
\cite{Love:Beamforming}. For the beam-nulling scheme, the optimal
ML receiver and the suboptimal MMSE receiver are used.

From Fig.~\ref{ber_bf_bn_4x4}, if the data rate is low, i.e.,
constellation size is low, beamforming outperforms beam-nulling.
If the data rate is high, i.e., constellation size is high,
beam-nulling outperforms beamforming at low and medium SNR,
however at high SNR beamforming outperforms beam-nulling. Also, as
can be seen, at the high data rate, even the beam-nulling scheme
with suboptimal MMSE receiver outperforms the beamforming scheme.

\subsection{Concatenation of Beam-nulling and LDC}

To further improve the performance of  beam-nulling with tractable
complexity, we propose to concatenate  beam-nulling with a linear
dispersion code. Note that to meet error-rate requirements,
multiple levels of error protection can be implemented. In this
study, we focus on space-time coding domain.

In this system, the information bits are first mapped into
symbols. The symbol stream is parsed into blocks of length
\(L=(N_t-1)T\). The symbol vector associated with one modulation
block is denoted by \(\mathbf{x}=[x_1, x_2, \ldots, x_L]^T\) with
$x_i \in \Omega\equiv\{\Omega_m | m=0, 1, \ldots, 2^\eta-1, \eta
\geq1\}$, i.e., a complex constellation of size \(2^\eta\), such
as \(2^\eta\)-QAM). The average symbol energy is assumed to be
\(1\), i.e., \(\frac{1}{2^\eta}\sum\limits_{m=0}^{2^\eta-1}
|\Omega_m|^2 =1\). Each symbol in a block will be mapped to a
dispersion matrix of size \(N_t \times T\) (i.e., $\mathbf{M}_i$)
and then combined linearly to form \((N_t-1)\) data streams over
\(T\) channel uses. The output \((N_t-1)\) data streams are
transmitted only over the subspace $\mathbf{\Phi}$ orthogonal to
the weakest spatial channel. The generation of the orthogonal
subspace $\mathbf{\Phi}$ is described in Section
\ref{subsec_BN_model}. The received signals can be written as
\begin{equation}\label{BN_LDC_rec_y}
\mathbf{y}=\sqrt{\frac{P}{N_t-1}}\mathbf{H}\mathbf{\Phi}\sum\limits_{i=1}^{L}\mathbf{M}_i
x_i+\mathbf{z}
\end{equation}
where $\mathbf{z}$ is additive white Gaussian noise vector with
\emph{i.i.d.} symmetrical complex Gaussian elements of zero mean
and variance $\sigma^{2}_{z}$. It is worthy to note that the
traditional beamforming scheme cannot work with space-time coding
since it can be viewed as a SISO channel. We compare the
concatenated scheme with the original schemes at the same data
rate.

\begin{figure}
\centering
\subfigure[R=3]{\includegraphics[width=3in]{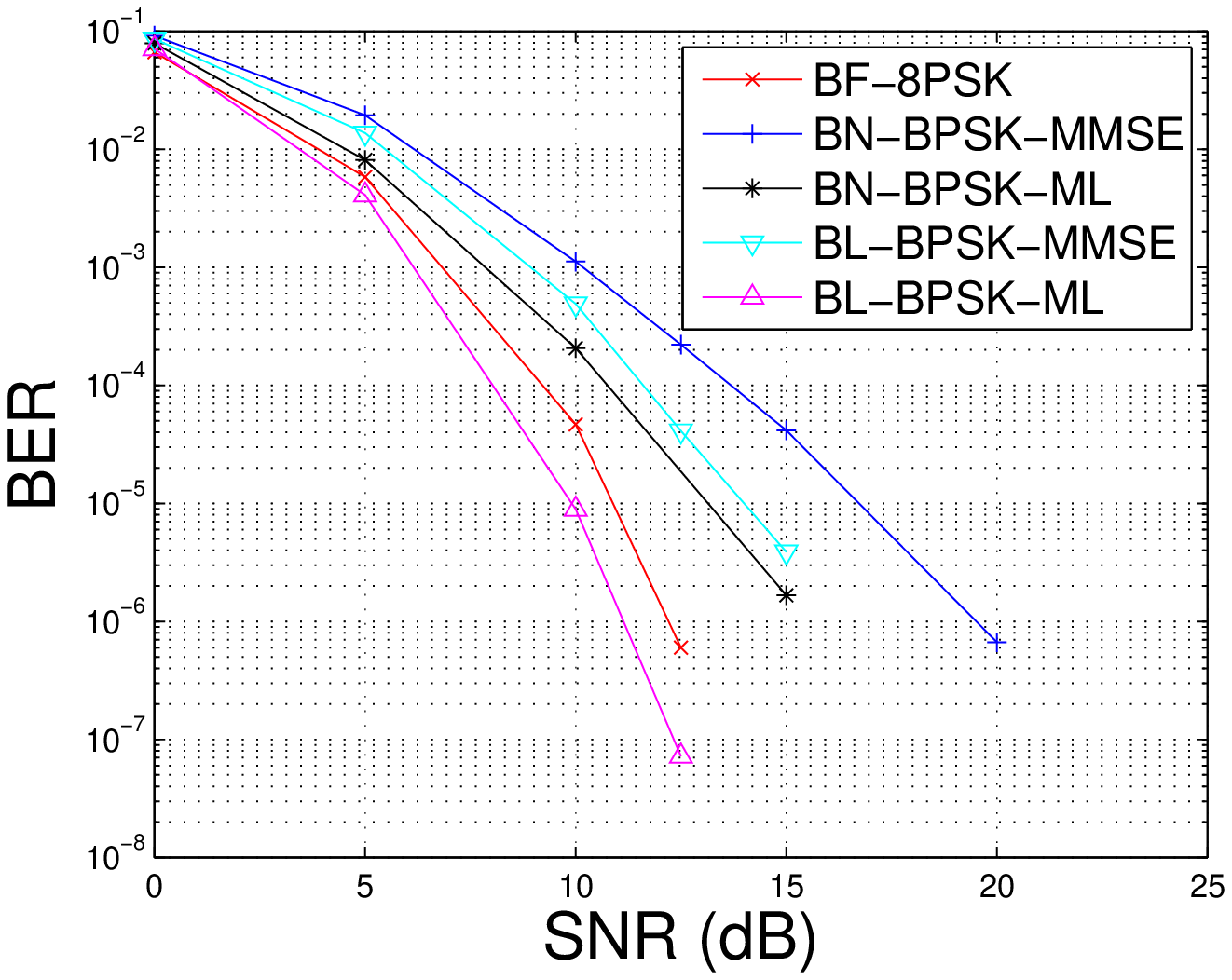}}
\hfil \subfigure[R=6]{\includegraphics[width=3in]
{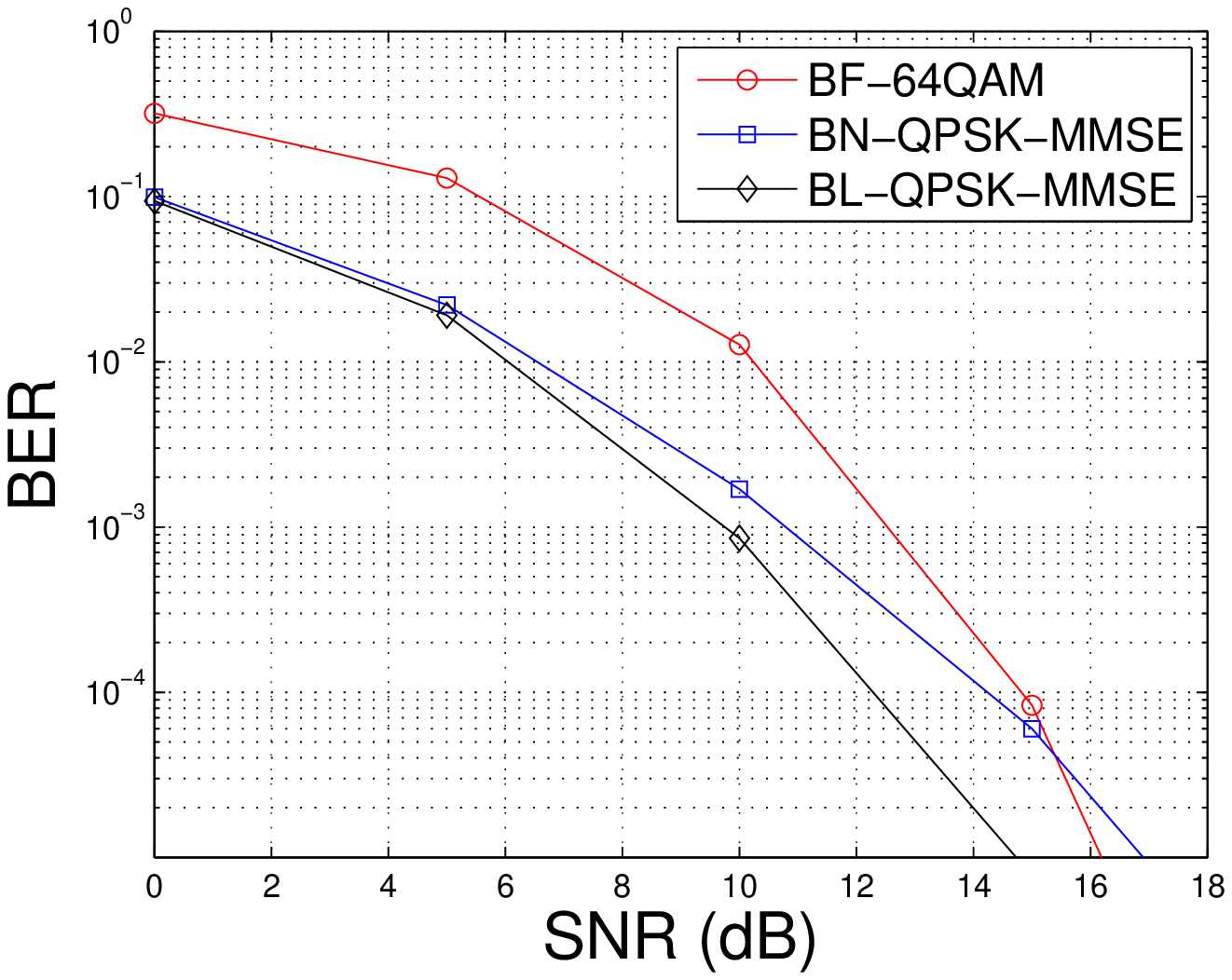}} \caption{\label{ber_bf_bn_bl_4x4}Comparison
over $4 \times 4$ Rayleigh fading channel.}
\end{figure}

In Fig.~\ref{ber_bf_bn_bl_4x4}, simulation results are compared
for various data rates $R$ over $4 \times 4$ Rayleigh flat fading
channels. In the figure, ``BL" denotes beam-nulling with LDC. As
can be seen, beam-nulling with LDC outperforms beam-nulling
without LDC using the same receiver. The performance of
beam-nulling with LDC using MMSE receiver is close to that of
beam-nulling without LDC using the optimal ML receiver.

Also it can be seen, if data rate is low, i.e., constellation size
is low, the performance of  beam-nulling with LDC can approach
that of beamforming at high SNR. If data rate is high, i.e.,
constellation size is high,  beam-nulling with LDC outperforms
 beamforming even when the suboptimal MMSE receiver is used.

\section{Extended Adaptive Frameworks}\label{sec_ext_frame}
For the beamforming and beam-nulling schemes, only one eigenvector
has been fed back to the transmitter. If more backward bandwidth
is available for feedback, e.g. $k$ eigenvectors, can be sent to
the transmitter for adaptation. With the feedback of $k$
eigenvectors, we can extend our frameworks, which will be called
multi-dimensional (MD) beamforming and MD beam-nulling. The
original schemes can be referred to as 1D-beamforming and
1D-beam-nulling. To save bandwidth, $k \leq
\lfloor\frac{N_t}{2}\rfloor$ should be satisfied, where
$\lfloor\cdot\rfloor$ denotes rounding towards minus infinity.
That is, whether the strongest or the weakest $k$ spatial
subchannels will be fed back according to the channel conditions.
For example, at low SNR, $k$ strongest spatial subchannels will be
fed back. At medium SNR, $k$ weakest spatial subchannels will be
fed back.

\subsection{MD Beamforming}
For MD beamforming, $\mathbf{v}_{1}, \ldots, \mathbf{v}_{k}$ are
fed back to the transmitter. $k$ symbols, saying
$\mathbf{x}_k=[x_1, x_2, \ldots, x_k]^T$, are transmitted. At the
receiver, the received vector can be written as
\begin{equation}\label{eigen_rec_y_k}
\mathbf{y}_k=\sqrt{\frac{P}{k}} \mathbf{H}[\mathbf{v}_{1} \ldots
\mathbf{v}_{k} ]\mathbf{x}_k+\mathbf{z}_k
\end{equation}
where $\mathbf{z}_k$ is the additive white Gaussian noise vector
with \emph{i.i.d.} symmetrical complex Gaussian elements of zero
mean and variance $\sigma^{2}_{z}$.

Consequently, the associated ergodic channel capacity can be found
as
\begin{equation}\label{cap_bf_k}
\bar{C}_{k,bf}=E\left[\sum\limits_{i=1}^{k}\log\left(1+\frac{P}{k~\sigma^{2}_{z}}\lambda_{i}^2\right)\right]
\end{equation}
Let $\rho = {P}/{\sigma^{2}_{z}}$ denote SNR. It is readily
checked that the capacity of MD beamforming is also concave and
monotonically increasing with respect to SNR $\rho$.
Differentiating the above ergodic capacity with respect to $\rho$,
we have
\begin{equation}
  \frac{\partial \bar{C}_{k,bf}}{\partial \rho} = E\left(\sum\limits_{i=1}^{k}\frac{1}{\rho+\frac{k}{\lambda_{i}^2}}\right)
\end{equation}
If $\rho \rightarrow 0$, equivalently at low SNR, it can be easily
found that
\begin{equation}
\frac{\partial \bar{C}_{(k-1),bf}}{\partial \rho} > \frac{\partial
\bar{C}_{k,bf}}{\partial \rho}, \rho \rightarrow 0
\end{equation}
If $\rho \rightarrow \infty$, equivalently at high SNR, it can be
easily found that
\begin{equation}
\frac{\partial \bar{C}_{k,bf}}{\partial \rho} > \frac{\partial
\bar{C}_{(k-1),bf}}{\partial \rho}, \rho \rightarrow \infty
\end{equation}
Note that $\bar{C}_{k,bf}=0$ for any $k$ when $\rho=0$ or minus
infinity in dB. Hence, at low SNR, the capacity of the $k$-D
beamforming scheme is worse than the $(k-1)$-D beamforming scheme
and while at high SNR, the capacity of the $k$-D beamforming
scheme is better than the $(k-1)$D beamforming scheme at the cost
of feedback bandwidth.

\subsection{MD Beam-nulling}
For MD beam-nulling, similar to $1$D beam-nulling, by a certain
rule, a subspace orthogonal to the $k$ weakest spatial channel is
constructed. That is, the following condition should be satisfied.
\begin{equation}\label{orth_subspace_k}
\mathbf{v}_{n}^H\mathbf{\Phi}^{(k)}=\mathbf{0}^T, \forall
n=N_t-k+1,\ldots, N_t.
\end{equation}
The $N_t \times (N_t-k)$ matrix $\mathbf{\Phi}^{(k)}=[\mathbf{g}_1
\mathbf{g}_2 \ldots \mathbf{g}_{N_t-k}]$ spans the
$(N_t-k)$-dimensional subspace.

At the transmitter, $N_t-k$ symbols denoted as $\mathbf{x}^{(k)}$
are transmitted only over the orthogonal subspace
$\mathbf{\Phi}^{(k)}$. The received signals at the receiver can be
written as
\begin{equation}\label{BN_rec_y_k}
\mathbf{y}^{(k)}=\sqrt{\frac{P}{N_t-k}}\mathbf{H}\mathbf{\Phi}^{(k)}\mathbf{x}^{(k)}+\mathbf{z}^{(k)}
\end{equation}
where $\mathbf{z}^{(k)}$ is additive white Gaussian noise vector
with \emph{i.i.d.} symmetrical complex Gaussian elements of zero
mean and variance $\sigma^{2}_{z}$. From (\ref{BN_rec_y_k}), the
associated instantaneous channel capacity with respect to
$\mathbf{H}$ can be found as
\begin{equation}\label{cap_bn_k}
\bar{C}_{bn}^{(k)}=E\left[\sum\limits_{i=1}^{N_t-k}\log\left(1+\frac{P}{(N_t-k)\sigma^{2}_{z}}\lambda_{i}^2\right)\right]
\end{equation}
It is readily checked that the capacity of MD beam-nulling is also
concave and monotonically increasing with respect to SNR $\rho$.
Let $\rho = {P}/{\sigma^{2}_{z}}$ denote SNR. Differentiating the
above ergodic capacity with respect to $\rho$, we have
\begin{equation}
  \frac{\partial \bar{C}^{(k)}_{bn}}{\partial \rho} = E\left(\sum\limits_{i=1}^{N_t-k}\frac{1}{\rho+\frac{N_t-k}{\lambda_{i}^2}}\right)
\end{equation}
If $\rho \rightarrow 0$, equivalently at low SNR, it can be easily
found that
\begin{equation}
\frac{\partial \bar{C}^{(k)}_{bn}}{\partial \rho} > \frac{\partial
\bar{C}^{(k-1)}_{bn}}{\partial \rho}, \rho \rightarrow 0
\end{equation}
If $\rho \rightarrow \infty$, equivalently at high SNR, it can be
easily found that
\begin{equation}
\frac{\partial \bar{C}^{(k-1)}_{bn}}{\partial \rho} >
\frac{\partial \bar{C}^{(k)}_{bn}} {\partial \rho}, \rho
\rightarrow \infty
\end{equation}
Note that $\bar{C}_{k,bn}=0$ for any $k$ when $\rho=0$ or minus
infinity in dB. Hence, at low SNR, the capacity of the $k$-D
beam-nulling scheme is better than the $(k-1)$-D beam-nulling
scheme at the cost of feedback bandwidth and while at high SNR,
the capacity of the $k$-D beam-nulling scheme is worse than the
$(k-1)$-D beam-nulling scheme.

\begin{figure}
\centering
\includegraphics[width=2.5in]{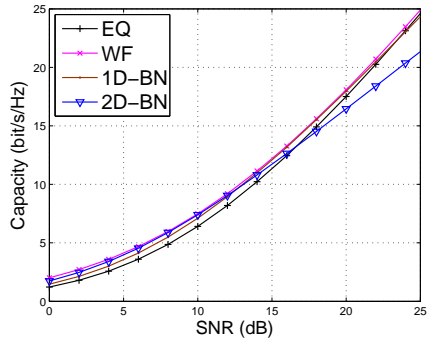}
\caption{ \label{md_bn_cap}MD beam-nulling over $5 \times 5$
Rayleigh fading channel.}
\end{figure}

For example, in Fig.~\ref{md_bn_cap}, capacities of 1D
beam-nulling and 2D beam-nulling schemes are compared with WF and
equal power scheme over $5 \times 5$ Rayleigh fading channel at
different SNR regions. At relatively low SNR, i.e., less than
13dB, the 2D beam-nulling scheme outperforms the 1D beam-nulling
scheme in terms of capacity at the price of feedback bandwidth.
While at relatively high SNR, i.e., more than 13dB, the
1D-beam-nulling scheme outperforms the 2D beam-nulling scheme as
predicted.

\subsection{Capacity Comparison of MD Schemes}
\begin{figure}
\centering
\includegraphics[width=2.5in]{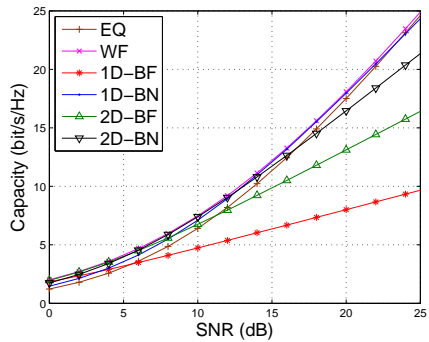}
\caption{\label{md_bf_bn_cap}Comparison over $5 \times 5$ Rayleigh
fading channel.}
\end{figure}

Here, over $5 \times 5$ Rayleigh fading channel, the MD schemes
are compared with WF and equal power schemes as shown in
Fig.~\ref{md_bf_bn_cap}. It can be readily check that, at
relatively low SNR, MD beamforming schemes are better than MD
beam-nulling schemes; while at relatively high SNR, the results
are opposite. Specifically, at very low SNR, i.e. less than 0dB,
the 1D beamforming scheme outperforms the other MD schemes. At the
SNR region between 0dB and 5.5dB, the 2D beamforming scheme
outperforms the other MD schemes. At the SNR region between 5.5dB
and 12.7dB, the 2D beam-nulling scheme outperforms the other MD
schemes. At the SNR region between 12.7dB and 23dB, the 1D
beam-nulling scheme outperforms the other MD schemes. Again, when
SNR is more than 23dB, the equal power scheme outperforms the
other suboptimal schemes.

\subsection{MD Schemes Concatenated with Linear Space-Time Code}

\begin{figure}
\centering
\includegraphics[width=2.5in]{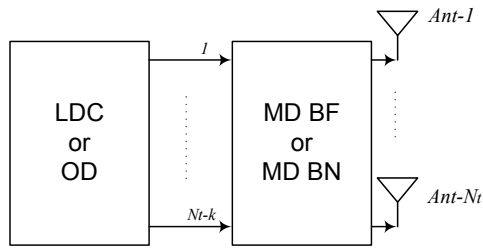}
\caption{\label{MD_scheme}Concatenated MD scheme.}
\end{figure}

MD beamforming scheme and MD beam-nulling scheme make $k$ and
$N_t-k$ spatial subchannels available, respectively. As a result,
they can concatenate with space-time schemes to improve
performance. For simplicity, space-time codes with linear
structure, such as high-rate LDCs \cite{Hassibi:LDC} and STBCs
\cite{Tarokh:STBC} (i.e., orthogonal design), are preferable. It
is worthy of noting that the 2D beamforming scheme in
\cite{Giannakis:Optimal-STBC-Channel-Mean} is just a special case
of MD beamforming. As shown in Fig.~\ref{MD_scheme}, we propose to
concatenate an MD scheme with an LDC or an STBC. In these figures
``OD" stands for orthogonal design.

Over $5 \times 5$ Rayleigh fading channel, concatenated MD schemes
are compared at various data rate. In the simulations, two
eigenvectors can be fed back to the transmitter. For an MD scheme
with LDC, a suboptimal linear MMSE receiver is applied. Since a MD
scheme with STBC are orthogonal, a matched filter is applied,
which is also optimal.

\begin{figure}
\centering
\includegraphics[width=2.5in]{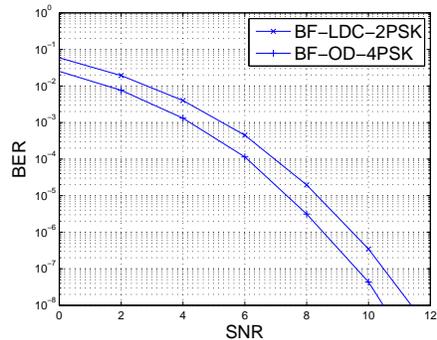}
\caption{\label{BER_MD_BF}BER of concatenated MD beamforming when
$R=2$.}
\end{figure}

\begin{figure}
\centering
\includegraphics[width=2.5in]{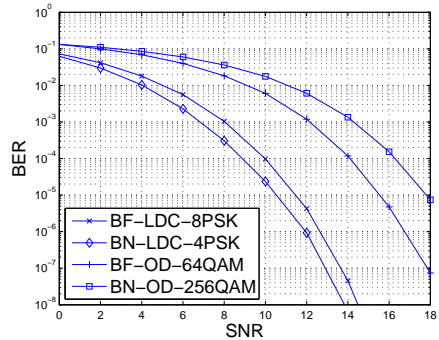}
\caption{\label{BER_MD_BF_BN}BER Comparison of concatenated MD
schemes when $R=6$.}
\end{figure}

In Fig.~\ref{BER_MD_BF}, MD beamforming scheme with STBC are
compared with MD beamforming scheme with LDC in terms of BER when
data rate is $R=2$. Also when $R=6$, Their BERs are shown in
Fig.~\ref{BER_MD_BF_BN}. From these figures, it is shown that at
high data rate, MD beamforming with LDC outperform MD beamforming
with STBC significantly even though a suboptimal MMSE receiver is
applied. Specifically, when BER is $10^{-5}$, the coding gain is
about $4$dB. At low data rate, MD beamforming with LDC performs
slightly worse than MD beamforming with STBC since the suboptimal
receiver is applied. Specifically, when BER is $10^{-5}$, the
coding gain is about $1$dB.

\begin{figure}
\centering
\includegraphics[width=2.5in]{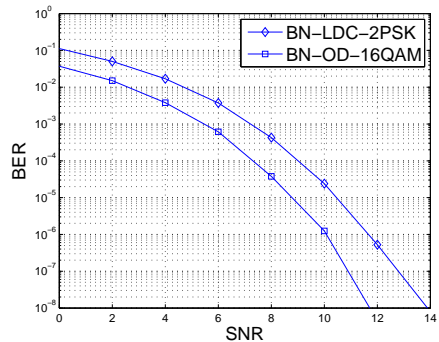}
\caption{\label{BER_MD_BN}BER of concatenated MD beam-nulling when
$R=3$.}
\end{figure}

In Fig.~\ref{BER_MD_BN}, MD beamforming scheme with STBC are
compared with MD beamforming scheme with LDC in terms of BER when
data rate is $R=3$. Also when $R=6$, Their BERs are shown in
Fig.~\ref{BER_MD_BF_BN}. From these figures, it is shown that at
high data rate, MD beam-nulling with LDC outperform MD
beam-nulling with STBC significantly even though a suboptimal MMSE
receiver is applied. Specifically, when BER is $10^{-5}$, the
coding gain is about $6.8$dB. At low data rate, MD beam-nulling
with LDC performs slightly worse than MD beam-nulling with STBC
since the suboptimal receiver is applied. Specifically, when BER
is $10^{-5}$, the coding gain is about $1.5$dB.

In Fig.~\ref{BER_MD_BF_BN}, four schemes are compared when data
rate is $R=6$. As shown in the figure, MD beam-nulling with LDC
has the best BER performance even suboptimal MMSE receiver is
used. In summary, MD scheme with LDC outperforms MD scheme with
STBC especially when the data rate is high. At low data rate, the
performance will depend on the receiver. At high data rate, MD
beam-nulling with LDC perform the best among the four schemes.

\section{Conclusions}\label{sec_con}

Based on the concept of spatial subchannels and inspired by the
beamforming scheme, we proposed a scheme called ``beam-nulling".
The new scheme exploits all spatial subchannels except the weakest
one and thus achieves significantly high capacity that approaches
the optimal water-filling scheme at medium signal-to-noise ratio.
The performance of beam-nulling with an MMSE receiver has been
analyzed and verified by numerical and simulation results. It has
been shown that if the data rate is low, beamforming outperforms
beam-nulling. If the data rate is high, beam-nulling outperforms
beamforming at low and medium SNR but beamforming outperforms at
high SNR. To achieve better performance and maintain tractable
complexity, beam-nulling was concatenated with a linear dispersion
code and it was demonstrated that if the data rate is low,
beam-nulling with a linear dispersion code can approach
beamforming at high SNR. If the data rate is high, beam-nulling
outperforms beamforming even with a suboptimal MMSE receiver. If
more than one eigenvector can be fed back to the transmitter, new
extended schemes based on the existing beamforming and the
proposed beam-nulling are proposed. The new schemes are called
multi-dimensional beamforming and multi-dimensional beam-nulling,
respectively. The theoretical analysis and numeric results in
terms of capacity are also provided to evaluate the new proposed
schemes. Both of MD schemes can be concatenated with an LDC or an
STBC. It is shown that the MD scheme with LDC can outperform the
MD scheme with STBC significantly when the data rate is high.
Additionally, at high data rate, MD beam-nulling with LDC
outperforms MD beamforming with LDC, MD beamforming with STBC and
MD beam-nulling with STBC.


\begin{thebibliography}{1}
\bibitem{Telatar:MIMOCapacity}
I.~E.~Telatar, ``Capacity of multi-antenna Gaussian channels,"
\emph{Eur. Trans. Telecom.}, vol 10, pp. 585-595, Nov. 1999.

\bibitem{Foschini:MIMOCapacity}
G.~J.~Foschini, M.~J.~Gans, ``On limits of wireless communications
in a fading environment when using multiple antennas,"
\emph{Wireless Personal Communications}, vol. 6, no. 3, pp.
311-335, 1998.

\bibitem{Cavers:VR_transmission}
J.~K.~Cavers, ``Variable-rate transmission for Rayleigh fading
channels," \emph{IEEE Transactions on Communications}, COM-20,
pp.15-22, 1972.

\bibitem{Goldsmith:VR_MQAM}
A.~J.~Goldsmith and S.-G.~Chua, ``Variable rate variable power
MQAM for fading channels," \emph{IEEE Trans. Commun.}, vol. 45,
no. 10, pp. 1218–1230, Oct. 1997.

\bibitem{Luo:Capacity_Time-Varying}
Z.~Luo, H.~Gao, Y.~Liu and J.~Gao ``Capacity Limits of
Time-Varying MIMO Channels,"  \emph{IEEE International Conference
On Communications} vol.2, pp. 795- 799, May 2005.

\bibitem{Cioffi:Iterative_WF}
W. Yu, W. Rhee, S. Boyd, and J. Cioffi, ``Iterative water-filling
for Gaussian vector multiple-access channels," \emph{IEEE Trans.
Inform. Theory}, vol.50, no. 1, pp. 145–152, Jan. 2004.

\bibitem{Goldsmith:Iterative_WF_sum}
N. Jindal, W. Rhee, S. Vishwanath, S. A. Jafar, and A. Goldsmith,
``Sum power iterative water-filling for multi-antenna gaussian
broadcast channels," \emph{IEEE Trans. Inform. Theory}, vol. 51,
no. 4, pp. 1570–1580, April 2005.

\bibitem{Ingram:Power_MIMO_Cap}
M. Demirkol and M. Ingram,  ``Power-controlled capacity for
interfering MIMO links," \emph{in Proc. IEEE Veh. Technol. Conf.
(VTC)}, Atlantic City, USA, Oct. 2001, pp. 187–191.


\bibitem{Shen:Comparison_Water-filling}
Z.~Shen, R. W.~Heath, Jr., J. G.~Andrews, and B. L.~Evans,
``Comparison of Space-Time Water-filling and Spatial Water-filling
for MIMO Fading Channels," \emph{in Proc. IEEE Int Global
Communications Conf.} vol. 1, pp. 431 – 435, Nov. 29-Dec. 3, 2004,
Dallas, TX, USA.

\bibitem{Luo:Design-adaptive}
Z.~Zhou and B.~Vucetic ``Design of adaptive modulation using
imperfect CSI in MIMO systems,"  \emph{2004 Eelectronics Letters}
 vol. 40 no. 17, Aug. 2004.

\bibitem{Zhang:QoS_WF}
X.~Zhang and B.~Ottersten, ``Power allocation and bit loading for
spatial multiplexing in MIMO systems," \emph{IEEE Int. Conf.on
Acoustics, Speech, and Signal Processing, 2003. Proceedings
(ICASSP '03)} vol.5 pp. 54-56, Apr. 2003.

\bibitem{Giannakis:Optimal-STBC-Channel-Mean}
S.~Zhou and G. B.~Giannakis, ``Optimal transmitter
eigen-beamforming and space-time block coding based on channel
mean feedback" \emph{IEEE Transactions on Signal Processing}, vol.
50, no. 10, October 2002.

\bibitem{Zhou:Channel_Prediction}
S.~Zhou and G.~B.~Giannakis, ``How accurate channel prediction
needs to be for transmit-beamforming with adaptive modulation over
Rayleigh MIMO channels," \emph{IEEE Trans. Wireless Comm.}, vol.
3, no. 4, pp. 1285 – 1294, July 2004.

\bibitem{Love:Beamforming}
D.~J.~Love, R.~W.~Heath,~Jr. and T.~Strohmer, ``Grassmannian
beamforming for multiple-input multiple-output wireless systems,"
\emph{IEEE Trans. Inform. Theory}, vol. 49, no. 10, pp. 2735 –
2747, Oct. 2003.

\bibitem{Zheng:Capacity_Precode}
J.~Zheng and B.~D.~Rao, ``Capacity analysis of MIMO systems using
limited feedback transmit precoding schemes," \emph{IEEE Trans. on
Signal Processing}, vol. 56, no. 7, pp. 2886 - 2901, July 2008.

\bibitem{Zhou:Adaptive_mod}
S.~Zhou and G.~B.~Giannakis, ``Adaptive modulation for
multiantenna transmissions with channel mean feedback," \emph{IEEE
Trans. Wireless Comm.}, vol.3, no.5, pp. 1626-1636, Sep. 2004.

\bibitem{Giannakis:Multiantenna-Beamforming-Constrained-Feedback}
P.~Xia and G.~B.~Giannakis, ``Multiantenna adaptive modulation
with beamforming based on bandwidth-constrained feedback,"
\emph{IEEE Transactions on Communications}, vol.53, no.3, March
2005.

\bibitem{Bishwarup:Performance-Beamforming-Quantized}
B.~Mondal and R.~W.~Heath, Jr., ``Performance analysis of
quantized beamforming MIMO systems," \emph{IEEE Transactions on
Signal Processing }, vol. 54, no. 12, Dec. 2006.

\bibitem{Goldsmith:BF_zero}
T. Yoo and A. Goldsmith, ``On the optimality of multiantenna
broadcast scheduling using zero-forcing beamforming," \emph{IEEE
J. Select. Areas in Commun.}, vol. 24, no. 3, pp. 528–541, March
2006.

\bibitem{Giannakis:BF_Quantifying}
S. Zhou, Z. Wang, and G. Giannakis, ``Quantifying the power loss
when transmit beamforming relies on finite rate feedback," IEEE
Trans. on Wireless Commun., vol. 4, no. 4, pp. 1948–1957, 2005.

\bibitem{Goldsmith:adaptive_BF_MO}
J. F. Paris and A. J. Goldsmith, ``Adaptive Modulation for MIMO
Beamforming under Average BER Constraints and Imperfect CSI,"
\emph{Proc. of Int. Conf. Comm.} ICC 2006, pp.1312-1317, June
2006.

\bibitem{Cavers:BF}
J.~K.~Cavers, ``Single-user and multiuser adaptive maximal ratio
transmission for Rayleigh channels," \emph{IEEE Trans. Veh.
Technol.}, vol. 49, no. 6, pp. 2043–2050, Nov. 2000.

\bibitem{Tarokh:STC_Perf_Const}
V.~Tarokh, N.~Seshadri, and A.~Calderbank, ``Space-time codes for
high data rate wireless communications: Performance criterion and
code construction," \emph{IEEE Trans. Inform. Theory}, vol. 44,
pp. 744-765, Mar. 1998.

\bibitem{Alamouti:STBC}
S.~Alamouti, ``A simple transmitter diversity scheme for wireless
communications," \emph{IEEE J. Select. Areas Commun.}, vol. 16,
pp. 1451-1458, Oct. 1998.

\bibitem{Tarokh:STBC}
V. Tarokh, H. Jafarkhani, and A. R. Calderbank, "Space-time block
code from orthogonal designs," \emph{IEEE Trans. Inform. Theory},
vol. 45, pp. 1456-1467, July 1999.

\bibitem{Hassibi:LDC} B.~Hassibi and
B.~Hochwald, ``High-rate codes that are linear in space and time,"
\emph{IEEE Trans. Inform. Theory}, vol. 48, pp. 1804-1824, July
2002.

\bibitem{Heath:LDC}
R.~W.~Heath and A.~Paulraj, ``Linear dispersion codes for MIMO
systems based on frame theory," \emph{IEEE Trans. on Signal
Processing}, vol. 50, No. 10, pp. 2429-2441, October 2002.

\bibitem{Ma:FDFR_STC}
X.~Ma and G.~B.~Giannakis, ``Full-diversity full-rate
complex-field space-time coding," \emph{IEEE Trans. Signal
Processing}, vol. 51, no. 11, pp. 2917-2930, July 2003.

\bibitem{Wu:Design4CSTM}
Z.~Wu and X.~F.~Wang, ``Design of coded space-time modulation,"
\emph{IEEE International Conference on Wireless Networks,
Communications and Mobile Computing}, vol. 2, pp. 1059-1064, Jun.
13-16, 2005.

\bibitem{Shannon:info_theory}
C.~E.~Shannon, ``A mathematical theory of communication",
\emph{Bell Syst. Tech. J.}, vol. 27, pp. 379–423 (Part one), pp.
623–656 (Part two), Oct. 1948, reprinted in book form, University
of Illinois Press, Urbana, 1949.

\bibitem{Lupas:LMUD} R.~Lupas and
S.~Verdu, ``Linear multiuser detectors for synchronous
code-division multiple-access channels," \emph{IEEE Trans. inform.
Theory}, vol. 35, pp. 123-136, Jan. 1989.

\bibitem{Poor:prob_MMSE}
H.~V.~Poor and S.~Verdu, ``Probability of error in MMSE multiuser
detection," \emph{IEEE Trans. inform. Theory}, vol. 43, pp.
858-871, May 1997.

\bibitem{Bohnke:SINR_Analysis}
R.~Bohnke and K.~Kammeyer, ``SINR Analysis for V-BLAST with
Ordered MMSE-SIC Detection," \emph{International Wireless
Communications and Mobile Computing Conference}, pp. 623-628, July
2006.

\bibitem{Proakis:DigComm}
J.~Proakis, \emph{Digital Communications}, 4th~ed. \hskip 1em plus
0.5em minus 0.4em\relax New York: McGraw-Hill, 2001.

%
%
%

\end{thebibliography}
\end{document}